# Scenario Planning and Nanotechnological Futures

Darryl Farber[1] and Akhlesh Lakhtakia[2,*]

[1]Science, Technology, and Society Program, Pennsylvania State University, University Park, PA 16802, USA (e-mail: dfarber@engr.psu.edu)

[2]Department of Engineering Science and Mechanics, Pennsylvania State University, University Park, PA 16802, USA (e-mail: akhlesh@psu.edu)

[*]Corresponding author

**Abstract.** Scenario planning may assist us in harnessing the benefits of nanotechnology and managing the associated risks for the good of the society. Scenario planning is a way to describe the present state of the world and develop several hypotheses about the future of the world, thereby enabling discussions about how the world ought to be. Scenario planning thus is not only a tool for learning and foresight, but also for leadership. Informed decision-making by experts and political leaders becomes possible, while simultaneously allaying public's perception of the risks of new and emerging technologies such as nanotechnology. Two scenarios of the societal impact of nanotechnology are the mixed-signals scenario and the confluence scenario. Technoscientists have major roles to play in both scenarios.

## 1. Introduction

Scenario planning is a way to identify and learn about the social, economic, and political factors that engender and influence socio-technical systems, and thus affect the adoption new technologies and their subsequent diffusion. Our focus in this paper is on nanotechnology. The societal challenges that nanotechnology presents constitute a specific case of the general challenges facing society from technological innovation, which is that the rate of technological innovation exceeds society's capacity (i) to understand the risks as well as the benefits of an innovation and (ii) to develop effective mechanisms to manage the risks. Our paper is written for undergraduate as well as graduate students in science and engineering, and may be useful for in-class discussions on the impact of science and engineering on societies.

*1.1 Nanotechnology*

Nanotechnology involves the use of materials in the length of scale of approximately 1-100-nm range in at least one of three spatial dimensions, in order to exploit phenomena or create and use structures devices and systems that have novel properties and functionalities in the 1-100-nm length scale [1]. Worldwide R&D funding for nanotechnology is estimated in multibillion dollars annually, with governments and private institutions spending roughly equally [2].

Nanotechnologies are classified [3] as incremental (exemplified by paints, cosmetics, thin films), evolutionary (nanotubes and quantum dots), and radical (e.g., molecular manufacturing), with the last class perhaps still in a pre-embryonic stage. The following topics provide virtually unlimited R&D opportunities: nanomaterials, nanometrology, integrated electronics and optoelectronics, nanomanufacturing, bionanotechnology, and nanomedicine. The scope of the nanotechnology-dependent economy extends over paints and chemicals, tools and dies, electronics, energy storage, textiles, medical drugs and products, cosmetics, sports equipment, and household consumables [4]. Commercially available nanotechnological products already include tennis rackets and balls, golf balls, bicycle frames, various medical and cosmetic emulsions, dental adhesives, and socks and footwarmers [5].

Desirable features for industrial application of any specific nanotechnology include (i) cost-effectiveness, particularly for newer classes of products; (ii) cradle-to-grave auditing of nanomaterials; and (iii) waste reduction. All three will continue to provide academic and industrial challenges for a while. The anticipated enhanced toxicity of nanomaterials, its potential impact on humans and other forms of life, and ways to cope effectively with it constitute another significant challenge [6].

A socioethical issue associated with the eventual ubiquity of nanotechnology is posed by new forms of intrusion into the privacy of the individual by private institutions as well as governments, including democratically elected governments. Creating governance regimes that manage this risk is needed. Additionally, because much of nanotechnology may be proprietary, certain private firms may be in a position to dominate markets, which has negative social welfare effects [6,7].

Given the rapid pace of nanotechnology R&D—and, even more so, given the symbiotic attributes of nanotechnology, biotechnology, information science and technology, and cognition science [8]—technoscientists, including physicists, must no longer dissociate themselves from the management of the associated risks for societies. Although economic, political, and legal decisions are generally made by non-technoscientists, these people need to be guided by technoscientists. Furthermore, the public, being the prime funder of nanotechnology R&D in one way or the other, has the right to demand greater transparency and accountability of taxpayer-funded research, which requires that technoscientists think through the broader implications of their research [7,9].

We also think that there is an ethical issue that technoscientists must begin to face. Principle 16 of the Rio Declaration on Environment and Development (often called the Polluter Pays Principle) [10] declares “that the polluter should, in principle, bear the cost of pollution, with due regard to the public interest and without distorting international trade and investment.” In the spirit of this principle, technoscientists may, in future, be held liable by societies suffering distress caused by even unintended by-products of technoscientific products and services.

Nanotechnoscientists, in particular, must reflect about the societal implications of their R&D activities while conducting those activities, guide their own R&D activities thereby, and advise economic, political, and legal decision-makers. In order to do so, the education of nanotechnoscientists must evolve to meaningfully include societal implications. We think the scenario-planning should be an integral part of technoscientific university-level education on the one hand and all technoscientific R&D activities on the other.

### *1.2 Scenarios*

The creation of a scenario is similar to modeling a physical, biological, or engineered system by a technoscientist. One difference however is that there is an interpretative element to social systems and to scenarios generated through the scenario-planning method. What we mean is that different people interpret factual data about the world differently, depending on the context of a situation as well as how a specific person comprehends the world. The elements of a scenario behave in ways not necessarily determined by physical laws or governed by invariant rules that are fully known to us; i.e., the elements of a scenario may have free will. For example, imagine a model of a car. One element of the model is the car's engine. Imagine further that whether the engine will turn on or not depends on the engine's decision that may be influenced by many factors, including if the engine likes the driver. An element's interpretation of the situation will influence how a social system will function, but may not necessarily be specifiable in advance because it depends on how that element interprets a specific situation.

A simple example will illustrate. If an automobile driver approaches an intersection and the traffic light is yellow, what will the driver do? Will the driver speed up or will the driver slow down? Even though the light is yellow, a driver's action at the moment of approaching an intersection is contextually specific. It may depend on the attitude of the driver (risk taker versus a risk-averse person), the time of day (rush hour versus early morning), and location (rural versus urban), among other factors. Different interpretations by different drivers imply different outcomes for the performance of the traffic system that intrinsically is unpredictable. What may be said with confidence is that different outcomes are possible for this complex situation.

### *1.3 Scenario Planning for Nanotechnological Advancement*

Scenario planning is a method to simplify the complexity and understand why different outcomes are possible. The use of the scenario-planning method is not to predict the future, but rather to learn more about how social, political, economic, and technological factors interact to create the current state of our world and how these factors may create future states of the world. It is a way to describe how the world *is* and then to develop hypotheses about how the world *may be*. Investigation of how the world is and consideration of how it may be together enable discussions about how the world *ought to be*.

With respect to nanotechnology, scenario planning may serve as a useful technique for scientists and engineers to engage with social scientists, humanists, and policy-makers in better understanding and reflecting about nanotechnology in society. Scenario planning may be an effective tool for creating a common language among different stakeholders and thereby bridging *two cultures*: that of science and technology on the one hand, and that of social studies and the humanities on the other.

## 2. What is Scenario Planning?

There are different conceptualizations of the term *scenario*. The conceptualization that we are discussing originated at the RAND Corporation in the 1960s and was further developed at Shell Corporation in the mid-1970s [11].

Scenario planning has its roots in systems analysis developed immediately after World War II at RAND [12, pp. 3-18]. As an extension of systems analysis, scenario planning was developed in response to the limitations of using probability in decision making. "Previously, analysts would have tried to optimize a single objective, or perhaps attempt to weigh different objectives and assess probabilities of outcomes in order to arrive at a best expected outcome" [12, p. 10]. Kahn, generally acknowledged as the father of scenario analysis, in his book *On Thermonuclear War* used the scenario-planning method to think through how a nuclear war may occur and to consider its consequences [13]. In a later book, entitled *The Year 2000,* Kahn and Weiner wrote [14]: "[The scenario-planning method is] especially valuable in the study and evaluation of the interaction of complex and/or uncertain factors." They continued [14]: "Scenarios are hypothetical sequences of events constructed for the purpose of focusing attention on causal processes and decision-points. They answer two kinds of questions: (1) Precisely how might some hypothetical situation come about, step by step? And (2) What alternatives exist, for an actor, at each step, for preventing, diverting, or facilitating the process." Educated at UCLA and Caltech as a physicist [15, pp. 63-65], Kahn had a clear understanding of the mathematical decision theory and its limits [12, p. 10]. From 1947 through 1961 Kahn worked at RAND as a physicist and during the period assisted Edward Teller and Hans Bethe on calculations for nuclear weapons [15, p. 64]. "In 1952 Kahn submitted his Monte Carlo studies to Cal Tech as a physics dissertation. But it was rejected on the grounds that academic research must not be commercially sponsored" [15, p. 65]. How have times changed! Kahn left RAND in 1961 and created the Hudson Institute, a think tank to explore broader issues facing society beyond defense. Kahn's Monte Carlo studies, nuclear physics, game theory, and systems analysis works are available at the RAND website [16], while his publications on scenario planning and the future are available at the Hudson Institute website [17].

Before Shell began to use scenario planning the company used a forecasting system known as Unified Planning Machinery (UPM) that provided detailed analysis for planning. According to Wack [11, p. 74], UPM was designed to reflect "a familiar, predictable world of 'more of the same.'" UPM looked out six years into the future. While using UPM in the mid-1960s, Shell also conducted a planning exercise that looked out to the year 2000 and as a result, it recognized that the assumption of a predictable oil market should be questioned and that a new planning method was needed.

Shell analysts addressed the creation of a method to provide insight on the possible ways the future may unfold, thereby improving decision-making abilities? Shell recognized that, to prepare better for distinctly different future business environments, it would need to "change the decision makers' assumptions about how the world works and compel them to reorganize their mental model of reality" [11, p. 74]. Through the process of creating and working with scenarios, one is forced to reflect on one's beliefs about how the world works.

The focus of Shell's scenarios was on the workings of the global oil markets and their evolution in time. In a book first published in 1991 with a new edition in 1998, Peter Schwartz, a former scenario planner at Shell, described the scenario-building process [18]. In general scenarios are created from the "inside-out" as opposed to "outside-in," which means one should initiate the creation of a scenario with a specific issue in mind and work towards identifying and clarifying factors in the larger world that may impact a specific concern rather than initially constructing global scenarios and then identifying how potential future worlds may influence a specific concern.

The first step in the scenario-building process then is the identification of a focal issue, which for us is the global governance of nanotechnology. The second step is the identification of elements influential in the relevant local environment: companies involved in nanotechnology research and development, potential products, potential consumers, and specific governmental regulations. In the third step, the scope is expanded to elements in the world at large and how those elements influence the local elements. For instance, the public's perception of risk could influence the demand for nanotechnological products as well as the extent of regulatory oversight. In the fourth step, these elements are ranked by importance and uncertainty. The most important and the most uncertain elements become the bases of different scenarios.

Some elements, being pre-determined, will appear in all of the different scenarios. Wack offered a clear way of thinking about the meaning of pre-determined elements in scenarios. "By pre-determined elements, I mean those events that have already occurred (or that almost certainly will occur but whose consequences have not yet unfolded). Suppose, for example, heavy monsoon rains hit the upper part of the Ganges River basin. With little doubt you know that something extraordinary will happened within two days at Rishikesh at the foothills of the Himalayas; in Allahabad, three or four days later; and at Benares, two days after that. You derive that knowledge not from gazing into a crystal ball but from simply recognizing the future implications of a rainfall that has already occurred" [11, p. 77].

With the elements of the scenarios identified, the next step is the development of the logic of every scenario. A scenario's logic is the rationale for the elements to fit together in that scenario: essentially, a story has to be woven from the elements. The story is effectively an explanation of why the elements of a scenario behave as they do. As Schwartz stated, "While in the end one may boil the logic down to the directions of a very few variables the process for getting there is not at all simple or mechanical. It is more like playing with a set of issues until you have reshaped and regrouped them in such a way that a logic emerges and a story can be told" [18, p. 244].

The scenario-building process is similar to the construction of a scientific theory. Scenarios are hypotheses of the future state of the world. They may also be thought of simulations of possible worlds. Scenario-planning forces coherent thought. Some critics claim that scenarios may be divorced from reality, and could therefore be misleading and even dangerous. Kahn provided an effective rebuttal to this criticism as follows [19]: "[O]ne must remember that the scenario is not used as a predictive device. The analyst is dealing with the unknown and to some degree unknowable future. … Imagination has always been one of the principal means for dealing in various ways with the future, and the scenario is simply one of many devices useful in stimulating and disciplining the imagination. To the extent that particular scenarios may be divorced from reality, the proper criticism would seem to be of particular scenarios rather than of the method. And of course unrealistic scenarios are often useful aids to discussion; if only to point out that the particular possibilities are unrealistic."

Scenarios though are not simple flights of fancy, imaginative exercises. Rather, they are an attempt to understand *the present* and to understand the factors that have shaped it. This understanding then may lead one to speculate how current trends may continue or how current trends may break and why. As Schwartz wrote [18, p. 9], "To operate in an uncertain world, people needed to be able to *reperceive*—to question their assumptions about the way the world works, so that they could see the world more clearly. The purpose of scenarios is to help yourself change your view of reality—to match it up more closely with reality as it is, and reality as it is going to be. *The end result, however, is not an accurate picture of tomorrow, but better decisions about the future.*"

## 3. The Design and Evolution of Nanotechnology-Risk Management Regimes

Although nanotechnology promises significant improvement in the well-being of humankind—through increased productivity in health care, environmental protection, and energy resources, to name a few areas—there is also concern regarding the potential environmental health and safety risks of the technology [7,20,21]. Lessard and Miller [22] defined risk as "the possibility that events, the resulting impacts, the associated actions, and the dynamic interactions among the three may turn out differently than anticipated." Scenarios then may be used to understand the possible causal sequence of how events may turn out differently than anticipated. Scenarios may also be used to better understand the design and evolution of a global nanotechnology risk-management regime. Regimes may be thought of as [23] "social institutions created to respond to the demand for governance relating to specific issues arising in a social setting that is anarchical in the sense that it lacks a centralized public authority or government in the ordinary meaning of the term." Speth and Haas [24, p. 83-84] stated that international regimes are "principles, norms, rules and decision-making procedures."

In the United States there is increasing recognition that environmental, health, and safety (EHS) regulations have not kept pace with innovation [25]. Although there is increasing concern about the EHS effects of nanotechnology, nanotechnology products are increasingly being allowed to enter the marketplace. The societal response to this challenge is to increase the amount of funding for EHS research. Approximately US$1.5 billion has been proposed in the US for nanotechnology research in fiscal year 2009, the share of EHS research being $76 million or about 5% [26].

The general problem that faces society from nanotechnological innovation is that, although engineers can increasingly create a variety of nanomaterials to perform specific functions with increased performance, understanding of the biological implications and movement through the biosphere remains primitive. Given the high diversity of engineered nanomaterials and the lack of standardized toxicological testing procedures [27,28,29], the central question is whether the societal EHS response is of commensurate scale to the challenge. Schierow recently provided a comprehensive review of the challenges and responses [30]. For example [30, p. 11], "A recent report by the Subcommittee on Science and Technology of FDA's Science Board concluded with respect to all FDA programs (again, not just nanotechnology) 'that science at the FDA is in a precarious position: the Agency suffers from serious scientific deficiencies and is *not positioned* (our emphasis) to meet current or emerging regulatory responsibilities.' " According to the Environmental Defense Fund (EDF) [31], the U.S. Environmental Protection Agency's (EPA) voluntary Nanoscale Materials Stewardship Program "will not deliver critically needed information and serves only to postpone key decisions on how best to mitigate nanotechnology's potential risks to human health and the environment".

The design and evolution of a nanotechnology risk-management regime will emerge from the interaction of stakeholders pursuing their self-interest and from the negotiations they will enter into to realize this interest. In a recent policy brief on nanotechnology [32], the International Risk Governance Council outlined a general approach to addressing risk and recommended the following actions:

(i) standardization of terminology and measurements;

(ii) development of worker guidelines to prevent risks, even if the risks are highly uncertain

(iii) better communication of risk to the public; and

(iv) increased funding for risk-related science.

Additionally, the Council recommended a global coordinated effort to:

(v) "ensure transparency of risk assessment data";

(vi) "synthesize and assess progress";

(vii) "consider the development of internationally compatible legally binding regulations for risk issues not amenable to voluntary restraints"; and

(viii) "make recommendations for further work."

Currently, a comprehensive legal regime to specifically regulate nanotechnology does not exist anywhere in the world, but recent scholarship recommends such an approach [33]. Governmental regulations on nanotechnology are formulated piecemeal, if at all [7, 34]. New nanotechnologies are evaluated on a case-by-case basis under existing regulations that had been framed to implement laws that either did not address or could not have addressed nanotechnology. In the United States, one way the Environmental Protection Agency may regulate a *new* chemical is under the Toxic Substances Control Act (TSCA) and a new pesticide under the Federal Insecticide, Fungicide, and Rodenticide Act (FIFRA). For instance, the EPA has recognized that nanoparticles may have unique hazards and encourages [35] "companies to discuss requirements for some specific nanoscale materials being used as pesticides. . . . [B]ecause nanoscale materials may have special properties, EPA's data requirements may need to be tailored to the specific characteristics of the product under consideration."

Benn and Westerhoff [36] recently demonstrated that commercially available nanosilver-containing socks release silver nanoparticles into the wastewater stream when they are washed. This release could affect subsequent bio-solid disposal as a fertilizer because of excess concentration of silver. Additionally, the silver nanoparticles may be detrimental to beneficial bacteria that removal ammonia from waste water [37]. In May 2008, the International Center for Technology Assessment [38] filed a legal petition with the EPA "demanding the agency use its pesticide regulation authority to stop the sale of numerous consumer products now using nanosized versions of silver. The legal action is the first challenge to EPA's failure to regulate nanomaterials." The EPA must respond to the petition within a "reasonable time," as reported in the New York Times [39].

The U.S. Food and Drug Agency (FDA) also has authority to regulate nanotechnology under existing regulations though it is not clear whether it has sufficient legal authority [40, 41].

Whether new national and/or international laws are needed in the short term or whether regulations based on existing laws will suffice is an issue to be debated. Regardless of the outcome, the anticipated ubiquity and pervasiveness of nanotechnology strongly enjoin the formulation of organizing principles in order to adequately assess the risks of nanotechnology. With their education and training in organizing technoscientific facts and procedures, technoscientists must be involved in that formulation.

## 4. Two Scenarios

Enunciation of possible futures of a risk-governance regime would definitely assist multiple stakeholders involved with nanotechnology to make informed decisions. Nanotechnology clearly promises a new industrial revolution; as such, there is a significant incentive on the part of business and governments to exploit this new technology for competitive advantage and economic growth [7]. Because of the great incentive for increasing research, development, demonstration, and deployment of nanotechnology, this element is considered pre-determined in our scenarios.

We also know that there is continued concern on EHS issues and funding for EHS research continues to increase [20,21,25]. Therefore, information about the potential hazards of nanotechnology will continue to emerge and grow. This is also a pre-determined element in our scenarios.

Although the outcomes of EHS research work are uncertain at this time, if we assume

(i) that some studies will be negative for hazardous effects and some studies will be positive, and

(ii) that nanomaterials in increasing amounts and novel types are being developed,

then we know that there will be an increase in the amount of information about the potential hazards of nanotechnology. Additionally, news of a potential hazard from nanotechnology will most likely receive greater attention from the media than news that a certain nanomaterial has been found to be safe, thereby potentially amplifying the public's perception of risk.

We do not know how society will interpret and respond to the increased amounts of negative information about nanotechnological hazards, even though there also will be increased amounts of positive information. Like the driver's interpretation of the yellow light, it is critically uncertain whether increased amounts of negative information about some nanotechnologies will increase the public's perception of risk and therefore increase the political pressure for regulatory action. The increase in public awareness will also have an affect on the response of industry. All scenarios thus must revolve around the flow and interpretation of increasing amounts of information on the risks of nanotechnology risk.

Let us now formulate two scenarios: the mixed-signals scenario and the confluence scenario.

*4.1 Mixed-Signals Scenario*

In this scenario, nanotechnological products were found to be both biologically benign and biologically hazardous in varying degrees, so much so that no clear cut conclusions emerges. It was difficult, using animal studies, to clearly isolate the effects from low doses of nanomaterials in humans; furthermore, the cumulative effect of low doses was difficult to determine because not enough data existed. Experts who urged a precautionary approach were countered by those experts who presume a nanomaterial is safe unless shown otherwise. The overall effect was that the public became confused.

Complicating matters was simply the flood of new toxicological studies done by many research groups, public as well as private, in many parts of the world. Even though researchers seemed to be progressing towards some general principles of nanotoxicity [42], it also appeared that the proliferation of novel nanomaterials leads to discoveries of heretofore unknown mechanisms of toxicity. Another problem was that different combinations of certain nanomaterials appeared to have different toxicological properties [43,44].

Although there was much discussion among the different stakeholders about the regulations needed, the framing of regulations had to depend on (i) the interpretation of data and experiments and (ii) whether the results obtained thereby constitute sound science. Questions about those nanotechnologies that could potentially directly and immediately affect human health—such as nanotechnology in food products, particularly products consumed by young children—generated the most public concern. Even though there was no consensus among the technoscientists, the public's perception of risk arose, and manufacturers of these nanotechnologies voluntarily stopped selling those nanomaterials.

An example of the mixed-signals scenario is furnished by the recent controversy over the safety of bisphenol-a (BPA). BPA is a chemical used in certain types of plastics. It is commonly found in bottles, tubing, and liners of food containers, among other consumer products. The FDA stated [45]: "Based on our ongoing review, we believe there is a large body of evidence that indicates that FDA-regulated products containing BPA currently on the market are safe and that exposure levels to BPA from food contact materials, including for infants and children, are below those that may cause health effects." However, "the National Toxicology Program (NTP), a federal interagency initiative, released a final report saying it has "some concern" that BPA is linked to health and developmental problems in humans. . . . Unfortunately, it is very difficult to offer advice on how the public should respond to this information," stated Michael Shelby, a director within a division of NTP [National Toxicology Program], in a statement [46]. "If parents are concerned, they can make the personal choice to reduce exposures of their infants and children to BPA. . . . Despite some mixed messages about BPA's safety, worries about the chemical have prompted some hefty action. Canada has said it plans on banning BPA in baby bottles; Wal-Mart Stores Inc., among other retailers, said in April it would stop selling baby bottles with the chemical; and more than 10 states have considered legislation banning the product in some baby and food products," the Wall Street Journal reported [46,47].

The main message of the mixed-signals scenario is that, because there is no agreed-upon framework for integrating new information and knowledge about nanotechnology, stakeholders will have difficulty in finding consensus about how to act, thus creating an adversarial environment about EHS implications.

### *4.2 Confluence Scenario*

In this scenario, it became increasingly recognized that what was once a trickle of new nanotechnology toxicological information over a few years to a decade would become a flood of new information. This insight was the key to organizing a global nanotechnological information infrastructure and knowledge base. It was also recognized that this infrastructure would have both designed features and evolving features. The capacity for new information to be integrated continuously into a matrix of existing knowledge in such a way that the capacity for recognizing emerging patterns of toxicity would also increase.

In 2008 there were already signs that such a global knowledge system was taking root. The International Alliance for NanoEHS Harmonization was one such effort [48]: it is "an interdisciplinary alliance of scientific experts, themselves currently active in all aspects of this arena, drawn from Europe, Japan and the United States that seeks to establish reproducible approaches for the study of nanoparticle hazards." To monitor the long-term environmental effects of nanoparticles, it was recognized that a nanotechnology component needed to be added to an Earth Observatory "movement." Different regional research groups from all over the world contributed to an effort to describe the human health and the natural environment in their region, such that collectively a detailed picture of the flows of energy and materials worldwide emerged. Although there was still technoscientific uncertainty, increasing amounts of information and knowledge were made intelligible so that a comprehensive, global, environmentally informed and risk-informed governance emerged, thereby enabling decision making based on informed consent at a level never before possible [49,50]. A comprehensive reform in education right from primary schools to the university level sustained the changes as well as the necessary mindset to accommodate the emergence of future unknowns [51,52].

The main message of the confluence scenario is that the provision of a framework allows the stakeholders to make informed decisions, thereby creating a cooperative environment about EHS implications. Furthermore, this scenario provides ongoing flexibility to decision makers because they can maintain a sustained and meaningful dialogue with the public and technoscientists, which is consistent with the US National Nanotechnology Initiative's strategy for nanotechnology-related EHS research [53, p. 8].

### *4.3 An Application for Nanotechnology*

The two scenarios described different societal responses to the issue of toxicity of nanoparticles [27,28,29]. Nanoparticles of a substance may be more toxic than larger particles of the same substance, because the enhanced surface area per unit volume of nanoparticulate matter generally makes it more chemically reactive. At the same time, nanoparticles could penetrate skin more easily than larger particles. However, as nanoparticles are expected to be used in small quantities––relative to, say, structural materials such as steel and concrete—the risk to general public by direct exposure is expected to be small in the near future. However, "the greatest potential for exposure … over the next few years will be in the workplace, both in industry and in universities," according to the Royal Society and the Royal Academy of Engineering [6, p. 42]. The risks involved include physical and biochemical problems following inhalation of nanoparticles, skin damage on penetration by nanoparticles and enhanced penetration of diseased skin, as well as explosions caused by spontaneous combustion of nanoparticles upon release in air.

The high toxicity of nanoparticles must be balanced by the possible therapeutic effects of medically administered nanoparticles [54]. Indeed, the enhanced reactivity due to high surface-to-volume ratio that makes nanoparticles toxic can also make them medically effective. Another benefit of nanoparticles is for enhancement of medical imagining to enable early diagnosis [55,56].

Carbon nanotubes are furnishing an excellent example of uncertainty. Pulmonary toxicity [57] as well as decreased cell function and oxidative stress [58] have been reported by some medical researchers. But other researchers have negated the finding of pulmonary toxicity, although nontoxic accumulation in the spleen were found [59]. Furthermore, evidence exists for the killing of breast-cancer cells by carbon nanotubes [54].

In the mixed-signals scenario, the public's perception of risk shall rise substantially. Because of perceived overwhelming harm that nanotubes are thought to cause, research-funding agencies will be under increased pressure to justify funding research on medical benefits of carbon nanotubes. Subsequently, researchers in their own interest will seek research opportunities in other areas because of the social stigma attached to carbon nanotubes. Additionally, fearful of potential lawsuits, manufacturers of carbon nanotubes shall stop producing carbon nanotubes for even non-medical purposes.

In the confluence scenario, because well-developed information and knowledge systems are in place, it is clear to the public what is known and what is unknown and to what degree. Even though there exist uncertainties, the public is generally confident that the regulatory agencies and manufacturers of nanotechnologies are working in a coordinated way such that the EHS issues are being such that the risk is acceptable. For example, not only are increasing amounts of toxicological, therapeutic, and diagnostic data on carbon nanotubes available though the Internet, the ability to interpret the significance of the results is also made more accessible through advances in the knowledge-and-learning-system technology, such as Web 2.0. Even though dosimetric research will be performed to determine safe levels of administration, and an official dosimetric consensus will emerge, much like the one that has emerged and continues to be refined for exposure to ionizing and non-ioninzing forms of radiation, there will be remaining questions about the chronic effects of very-low doses of exposures as well as interaction effects of multiple toxic agents at low dose.

## 5. The Education of Nanotechnoscientists

Education at the pre-university and the university levels is mostly of the *just-in-case* (JIC) variety. Certain topics are taught just in case those topics turn out to be useful to the students in later years. As the future cannot be accurately predicted, it is best that students acquire a broad background and many skills. One of us (AL) has argued elsewhere [51] that successful education in nanotechnology would require supplementation, not replacement, by a different instructional approach: *just-in-time* (JIT) education. This supplementary approach was enunciated in 1992 [6], drawing upon the principles of JIT manufacturing [61]. Today, it is heavily applied in the information technology and distance-education sectors for training and retraining the adult workforce, as googling will readily prove. It is also used to enhance learning in heavily subscribed lower-division undergraduate courses at US universities [62].

The education of nanotechnoscientists (as well as of other types of technoscientists) in scenario planning has features of both the JIC and the JIT varieties. Classification as JIC education is justified by the unpredictability of the future. Students must be prepared to formulate several scenarios for the adoption of a new nanotechnological process or the manufacture and sale of a new nanotechnology-based product. Classification as JIT education is justified by the diversity of resources that will be needed to quickly formulate several scenarios.

We envision education in scenario planning to comprise both in-class discussions and homework projects. In-class discussions can be focused on planning scenarios for emerging nanotechnologies such as quantum dots and metallic nanoparticles. Enough has been written about the benefits and risks of these nanotechnologies that several different scenarios of public discourse and political, legislative, and economic activity can be brainstormed. Homework projects could deal with scenario planning for the outcomes of the research projects of particular graduate students.

## 6. Concluding Remarks

Good scenarios challenge our ways of thinking about the world and the use of scenario analysis for better understanding of the social and ethical implications of nanotechnology is acknowledged as such in the National Nanotechnology Initiative Strategic Plan [63, p. 31]. Arie De Geus, a former Shell scenario planning has stated [64, p. 46]: “[Scenarios] are tools for foresight—discussions and documents whose purpose is not a prediction or a plan, but a change in the mind-set of the people who use them. By telling stories about the future in the context of our own perceptions of the present, we open our eyes for developments which in the normal course of daily life are indeed 'unthinkable'."

Thinking through the future via either *Mixed Signals* or *Confluence* may provide the decision makers a way of recognizing the terrain they currently inhabit so that they may craft strategies that are robust for the world that may unfold resembling either scenario. In these scenarios, a methodology to integrate new EHS-relevant information and knowledge about nanotechnology turns out to be an important condition that potentially may influence the rate of nanotechnological innovation because it may influence whether stakeholders are adversarial or cooperative in their interactions. Scenario planning combined with systems analysis thus is not only a tool for learning and foresight, but also for developing a philosophy of nanotechnology that fosters ethical leadership [65].

The centrality of technoscientific information and knowledge in both scenarios presented here, as well as in other possible scenarios, indicates that technoscientists—by attempting to quantify uncertainties—will play a major role towards informed-decision-making by political and business leaders. Technoscientists must understand that the rate of technological innovation exceeds society's capacity to cope with innovation, accept their responsibilities to consider the possible societal impacts of nanotechnology (and other socially transformative technologies), keep the public informed of both benefits and risks thereof, and prepare future generations of technoscientists who are socially aware and good citizens.

**References**

[1] Roco M C and Bainbridge W S (eds.) 2001 *Societal Implications of Nanoscience and Nanotechnology* (Arlington, VA, USA: National Science Foundation)

[2] Lux Research, Inc. 2004 *The Nanotech Report 2004. Investment Overview and Market Research for Nanotechnology* (New York, NY, USA: Lux Research)

[3] Jones R 2004 The future of nanotechnology *Physics World* **17**(8) 25-9

[4] Meridian Institute 2005 *Nanotechnology and the Poor: Opportunities and Risk* (Washington, DC, USA: Meridian Institute); available at: http://www.nanoandthepoor.org

[5] United Nations Educational, Scientific and Cultural Organization 2006 *The Ethics and Politics of Nanotechnology* (Paris, France: UNESCO); available at: http://www.unesco.org/shs/ethics

[6] Royal Society and Royal Academy of Engineering 2004 *Nanoscience and Nanotechnologies: Opportunities and Uncertainties* (London, United Kingdom: Royal Society)

[7] Munshi D, Kurian P, Bartlett R V and Lakhtakia A 2007 A map of the nanoworld: Sizing up the science, politics, and business of the infinitesimal *Futures* **39** 432-52

[8] Roco M C and Bainbridge W S (eds.) 2002 *Converging Technologies for Improving Human Performance: Nanotechnology, Biotechnology, Information Technology and Cognitive*

*Science* (Arlington, VA, USA: National Science Foundation)

[9] Rogers-Hayden T and Pidgeon N 2007 Moving engagement "upstream"? Nanotechnologies and the Royal Society and Royal Academy of Engineering's inquiry *Public Understand Sci.* **16** 345-64

[10] United Nations Environment Program 2000 *GEO-2000 Economic Instruments*; available at http://www.unep.org/Geo2000/english/0138.htm

[11] Wack P 1985 Scenarios: Uncharted waters ahead *Harvard Business Review* (September-October) 73-89

[12] Cooke R M 1991 *Experts in Uncertainty: Opinion and Subjective Probability in Science* (Oxford, United Kingdom: Oxford University Press)

[13] Kahn H 1960 *On Thermonuclear War* (Princeton, NJ, USA: Princeton University Press)

[14] Kahn H and Weiner A J 1967 *The Year 2000: A Framework for Speculation on the Next Thirty-Three Years* (New York, NY, USA: MacMillan) p. 6

[15] Ghamari-Tabrizi S 2005 *The Worlds of Herman Kahn: The Intuitive Science of Thermonuclear War* (Cambridge, MA, USA: Harvard University Press)

[16] RAND http://www.rand.org/pubs/authors/k/kahn_herman.html

[17] Hudson Institute http://www.hudson.org/index.cfm?fuseaction=HermanKahn

[18] Schwartz P 1998 *The Art of the Long View: Planning for the Future in an Uncertain World* (Chichester, United Kingdom: Wiley) pp. 241-248

[19] Kahn H 1990 *The Use of Scenarios* (Washington, DC, USA: Hudson Institute); available at: http://www.hudson.org/index.cfm?fuseaction=publication_details&id=2214 (This article was abstracted from pp. 262-264 of Ref. 4.)

[20] Lloyd's of London 2007 *Nanotechnology Recent Developments, Risks, and Opportunities* available at

http://www.lloyds.com/Lloyds_Market/Tools_and_reference/Exposure_Management/Emerging__risks/Emerging_risks.htm

[21] Environmental Defense Fund and Dupont 2008 *Nanorisk Framework* available at: http://www.nanoriskframework.com/page.cfm?tagID=1095

[22] Lessard D and Miller R 2000 Mapping and facing the landscape of risks. In: Miller R and Lessard D (eds.) 2000 *The Strategic Management of Large Engineering Projects: Shaping Institutions, Risks, and Governance* (Cambridge, MA, USA: MIT Press) p. 76

[23] Breitmeier H, Young O R and Zurn M 2006 *Analyzing International Environmental Regimes: From Case Study to Database* (Cambridge, MA, USA: MIT Press) p. 3

[24] Speth J G and Haas P M 2006 *Global Environmental Governance* (Washington, DC, USA: Island Press)

[25] Davies J C 2008 *Nanotechnology Oversight: An Agenda for the New Administration* (Washington, DC, USA: Woodrow Wilson International Center for Scholars)

[26] National Nanotechnology Initiative 2008 *NNI FY09 Budget* available at http://www.nano.gov/html/news/releases/20080214_NNI_Releases_FY09_Budget_Highlights.html

[27] Hoet P H M, Brüske-Hohlfield I and Salata O V 2004 Nanoparticles – known and unknown health risks *J. Nanobiotechnol.* **2** 12

[28] Hurt R H, Monthioux M and Kane A 2006 Toxicology of carbon nanomaterials: Status, trends, and perspectives on the special issue Carbon **44** 1028-33

[29] Dobrovolskaia M A and McNeil S E 2007 Immunological properties of engineered nanomaterials *Nature Nanotechnol.* **2** 469-78

[30] Schierow L-J 2008 *Engineered Nanoscale Materials and Derivative Products: Regulatory Challenges (RL34332)* (Washington, DC, USA: Congressional Research Service); available at: http://opencrs.com/document/RL34332

[31] Environmental Defense Fund 2008 EPA's voluntary program for nanomaterials still too little, too late; available at: http://www.edf.org/pressrelease.cfm?contentID=7564

[32] International Risk Governance Council 2007 *Nanotechnology Risk Governance* (Geneva, Switzerland: IRGC); available at: http://www.irgc.org

[33] Abbott K W, Marchant G E and Sylvester D J 2007 A framework convention for nanotechnology (Working Paper #2), Center for the Study of Law, Science, and Technology, Arizona State University, Tempe, AZ, USA

[34] Bowman D M and Hodge G A 2006 Nanotechnology: Mapping the wild regulatory frontier *Futures* **38** 1060-73

[35] US Environmental Protection Agency 2008 Pesticide issues in the works: nanotechnology, the science of the small; available at: http://www.epa.gov/pesticides/about/intheworks/nanotechnology.htm

[36] Benn T M and Westerhoff P 2008 Nanoparticle silver released into water from commercially available sock fabrics *Environ. Sci. Technol*. **42** 4133-39.

[37] Science Daily 2008 Silver nanoparticles may be killing beneficial bacteria in wastewater treatment; available at: http://www.sciencedaily.com/releases/2008/04/080429135502.htm

[38] International Center for Technology Assessment 2008 Legal actions: Nanotechnology; available at: http://www.icta.org/global/actions.cfm?page=15&type=364&topic=8

[39] Feder B 2008 No silver bullets *New York Times* May 6

[40] Taylor M R 2006 *Regulatory the Products of Nanotechnology: Does FDA Have the Tools it Needs?* (Washington, DC, USA: Woodrow Wilson International Center for Scholars)

[41] US Food and Drug Administration 2008 FDA regulation of nanotechnology products; available at: http://www.fda.gov/nanotechnology/regulation.html

[42] Service R F 2008 Can high-speed tests sort out which nanomaterials are safe? *Science* **321** 1036-7

[43] US Environmental Protection Agency 2003 A Framework for a Computational Toxicology Research Program in ORD (EPA/600/R-03/065) (Washington, DC, USA: Government Printing Office)

[44] Rabinowitz J R, Goldsmith M-R, Little S B and Pasquinelli M A 2008 Computational molecular modeling for evaluating the toxicity of environmental chemicals: Prioritizing bioassay requirements *Environ. Health Perspect.* **116** 573-7

[45] US Food and Drug Administration 2008 Bisphenol A (BPA); available at: http://www.fda.gov/oc/opacom/hottopics/bpa.html

[46] Favole J 2008 Federal health experts remain concerned about BPA safety *Wall Street Journal Health Blog* September 3; available at: http://blogs.wsj.com/health/2008/09/03/federal-health-experts-remain-concerned-about-bpa-safety/

[47] Favole J A and Mundy A 2008. FDA seeks data on medical devices, drugs containing BPA. *Wall Street Journal* October 15.

[48] International Alliance for NanoEHS Harmonization 2008 http://nanoehsalliance.org/

[49] Kleindorfer P R and Orts E W. 1998 Informational regulation of environmental risks *Risk Analysis* **18** 155-70

[50] Shrader-Frechette K 2007 Nanotoxicology and ethical conditions for informed *Nanoethics* **1** 47-56.

[51] Lakhtakia A 2006 Priming pre-university education for nanotechnology *Curr. Sci.* **90** 37-40

[52] Loftus M 2006 All things great and small *Prism* **15**(8); available at: http://www.prism-magazine.org/apr06/tt_01.cfm

[53] National Science and Technology Council 2008 Strategy for nanotechnology-related environmental, health, and safety research; available at: http://www.nano.gov/html/society/EHS.html

[54] Panchapakesan B, Lu S, Sivakumar K, Teker K, Cesarone G and Wickstrom E 2005 Single-wall carbon nanotube nanobomb agents for killing breast cancer cells *NanoBiotechnol.* **1** 133-8

[55] Loo C, Lin A, Hirsch L, Lee M-H, Barton J, Halas N, West J and Drezek R 2004 Nanoshell-enabled photonics-based imaging and therapy of cancer *Technol. Cancer Res. Treat.* **3** 33-

[56] Nammalavar V, Wang A M and Drezek R A 2007 Enhanced gold nanoshell scattering contrast using angled fiber probes *J. Nanophoton.* **1** 013510

[57] Lam C W, James J T, McCluskey R and Hunter R L 2004 Pulmonary toxicity of single-wall carbon nanotubes in mice 7 and 90 days after intratracheal instillation *Toxicol. Sci.* **77** 3-5.

[58] Mitchell L A, Gao J, Vander Wal R, Gigliotti A, Burchiel S W and McDonald J D 2007 Pulmonary and systemic immune response to inhaled multiwalled carbon nanotubes *Toxicol. Sci.* **100** 203-14

[59] Schipper M L, Nakayama-Ratchford N, Davis C R, Kam N W S, Chu P, Liu Z, Sun X, Dai H and Gambhir S S 2008 A pilot toxicology study of single-walled carbon nanotubes in a small sample of mice *Nature Nanotechnol.* **3** 216-21

[60] Hudspeth D 1992 Just-in-time education *Educ. Technol.* **32**(6) 7-11

[61] Ohno T 1988 *Toyota Production System: Beyond Large-Scale Production* (New York, NY, USA: Productivity Press)

[62] Novak G M, Patterson E T, Gavrin A and Enger R C 1998 Just-in-time teaching: Active learner pedagogy with www; available at: http://webphysics.iupui.edu/JITT/ccjitt.html

[63] National Science and Technology Council 2007 The National Nanotechnology Initiative strategic plan; available at http://www.nano.gov

[64] De Greus, A 1997 *The Living Company* (Cambridge, MA, USA: Harvard Business School Press) ch 3

[65] Farber D, Pietrucha M T and Lakhtakia A 2008 Systems and scenarios for a philosophy of engineering, *Interdisciplinary Sci. Rev.* **33** xxx-xxx (accepted for publication).